\documentstyle[aps]{revtex}
\input{epsf}

\begin{document}

\title{Diffusive growth of polydisperse hard-sphere crystals}
\author{R. M. L. Evans and C. B. Holmes \\
{\it Department of Physics and Astronomy, The University of
Edinburgh, Edinburgh EH9 3JZ, U.K.}}

\date{February 21, 2001}

\maketitle

\begin{abstract} Unlike atoms, colloidal particles are not identical,
but can only be synthesised within a finite size tolerance.  Colloids
are therefore polydisperse, i.e.~mixtures of infinitely many
components with sizes drawn from a continuous distribution. We model
the crystallisation of hard-sphere colloids (with/without attractions)
from an initially amorphous phase. Though the polydisperse hard-sphere
phase diagram has been widely studied, it is not straightforwardly
applicable to real colloidal crystals, since they are inevitably out
of equilibrium.  The process by which colloidal crystals form
determines the size distribution of the particles that comprise
them. Once frozen into the crystal lattice, the particles are caged so
that the composition cannot subsequently relax to the equilibrium
optimum. We predict that the mean size of colloidal particles
incorporated into a crystal is smaller than anticipated by equilibrium
calculations. This is because small particles diffuse fastest and
therefore arrive at the crystal in disproportionate abundance.
\end{abstract}

\vspace{-5mm}

% 64.75.+g 	Solubility, segregation, and mixing; phase separation
% 81.10.A	Nucleation, crystal growth
% 82.70.Dd      Colloids

\section{Introduction}
\label{intro}

How many ways are there to pack balls into a box? The answer, of
course, is infinitely many, since each ball can be placed in a
continuum of different positions. However, if one tried to count the
different arrangements, one would notice that many of them are very
similar. The predominant packing arrangements for a given
concentration $\phi$ (the fraction of the total volume occupied by the
balls) are summarised by the hard-sphere equilibrium phase diagram,
which is well known when all spheres are of equal size
\cite{Hoover68}. A face-centred-cubic (FCC) crystal phase exists for
any concentration from close packing at $\phi=\pi/\sqrt{18}$
\cite{Hales} down to $\phi=0.545$. For $\phi<0.494$ a fluid phase
exists in which configurations are amorphous, and between these two
concentrations, statistical weight is dominated by systems partitioned
into fluid and crystalline regions. On the other hand, when
differently-sized balls are considered, the full phase diagram is not
known. It may seem at first surprising that such a straightforward
question is still a field of active research.

The subject is not merely of idle interest. Many real substances,
particularly colloids, are composed of spherical particles with
negligible energetic interactions except for a hard repulsion at
contact. For a further, very large class of substances, the
hard-sphere model is a useful starting point for more accurate
theories \cite{WCA}.  Studying their configurations is of central
importance to our understanding of matter, and has implications for
the development of new materials.

It is often assumed that the configurations observed in a real
material, such as a hard-sphere colloid, are those of highest entropy
(lowest free energy for non-hard-sphere interactions), i.e.~those
which dominate the counting of all possible arrangements. On this
equilibrium assumption, many efforts have been devoted to exploring
the elaborate phase diagram of hard-sphere mixtures. Since the
particles never exchange potential energy, temperature does not
influence their configurations, only their speeds. So the phase
diagram has only as many axes as there are different sizes of sphere
in the mixture, each axis representing the concentration of one type
of particle. When just two species of balls are mixed, the diagram is
already complicated, with theory
\cite{Eldridge95} and experiment \cite{Bartlett92} finding regions of
fluid, pure crystals, and crystalline alloys of the A and B type
particles in the ratios AB, AB$_2$ or even AB$_{13}$ depending on the
size ratio. In the `polydisperse' hard-sphere phase diagram, for
systems such as real colloids, in which each particle is slightly
different from every other, an infinite number of axes is needed to
span the space of all possible compositions. A small corner of this
phase diagram has been charted by experiments \cite{Davesthesis},
simulations \cite{Bolhuis96} and theories based on phenomenological
free energies \cite{Bartlett97}, integral equations \cite{Barrat86},
cell theory \cite{Sear99} and perturbation about a monodisperse system
\cite{Evans98,Evans2001}. When the overall concentration of particles
is low, an amorphous fluid phase inevitably results. At higher
concentrations, it is expected that crystals of quite monodisperse
(similarly sized) particles can form, even when the overall size
distribution is broad \cite{Bolhuis96}, in which case, particles of
the wrong size to be included in the crystal remain in a coexisting
fluid phase. Many predictions exist for the equilibrium distribution
of particle sizes in a polydisperse crystal at coexistence with a
fluid.

We shall argue that such equilibrium states are not as
straightforwardly applicable to colloidal systems as might be
expected, and shall calculate an alternative distribution of particle
sizes, which we expect to comprise the crystallites observed in real
suspensions. The size distribution of particles incorporated into
colloidal crystals is determined by the kinetics of their formation
and, as we shall explain, never relaxes to the equilibrium
distribution, remaining instead as a relic of that process. Despite
the fact that the state of the system evolves while the crystals are
growing, we shall show that a large temporal regime exists during
which the distribution of species incorporated into a growing crystal
is invariant with time. Hence, during this regime, there exists a
unique solution for the distribution constituting the crystal. We
expect this kinetically mediated distribution to be trapped within
colloidal crystallites in laboratory samples that are quiescent and
therefore {\em appear} to have reach equilibrium. We shall find that
the particles buried deep within a crystallite are on average {\em
smaller} than would be expected from an equilibrium calculation. If
attractions exist between the hard particles, the effect can be so
marked as to swamp the thermodynamic driving force that favours {\em
larger} than average particles in the equilibrium crystal phase.

The rest of the article is organised as follows. In the next section
we shall discuss the formal procedure for obtaining the equilibrium
phase diagram of an $n-$component hard sphere system. We then go on to
discuss the evidence that experimental hard sphere systems are not at
equilibrium. In section \ref{GenDiffusion} we construct the most
general equations of motion for diffusive growth of a polydisperse
hard-sphere crystal, and show that there exists a time regime during
which the distribution of particles dynamically incorporated into the
crystal has a time-invariant solution. It will transpire that the
crystalline distribution is one that appears on the phase boundary in
the {\em equilibrium} phase diagram, but is not located on the
expected tie line, due to temporary violation of the lever rule. Using
a low-concentration approximation for the diffusion coefficients in
the fluid phase, we shall derive, in section \ref{ModLever}, a simple
replacement for the lever rule, which yields the appropriate
distributions for this non-equilibrium situation. In section
\ref{Perturb} we derive the conditions for local mechanical and
chemical balance at the crystal-fluid interface (equivalent to finding
binodals in the equilibrium phase diagram) using, for convenience, a
perturbative approximation that is valid for narrow
distributions. Section \ref{results} contains our results, and we
conclude in section \ref{conc}.

\section{Equilibrium Phase Diagrams}

As explained above, the phase diagram of single sized (monodisperse)
hard spheres is well known. The phase behaviour may be calculated via
a theoretical treatment in the following way. First, an expression for
the Helmholtz free energy of the fluid and crystalline phases must be
found. From this, one has knowledge of all thermodynamic quantities.
The phase boundaries are then determined by finding the values of
$\phi$ for which the pressure and chemical potential are equal in two
phases, signalling mechanical and chemical equilibrium between the
phases. The volumes of the two coexisting phases may be found by
application of the lever rule,
\begin{equation}
\label{lever}
	\phi_c V_c + \phi_f V_f = \phi_p V
\end{equation}
in terms of the system volume $V=V_c+V_f$.  The subscript $c$ refers
to a quantity in the crystal phase, $f$ refers to the fluid phase, and
$p$ to the overall, or {\it parent} distribution. The lever rule
follows from conservation of material in the system.

The situation is immediately complicated if we consider a system
containing differently sized particles, although the formal procedure
is wholly analogous. We illustrate this firstly by a glance at a
system of binary hard spheres. The phase diagram is now 2-dimensional,
so the task of finding coexisting phases is a problem in 4 variables,
because one must determine the concentrations of 2 species for each of
the 2 phases. Coexisting phases must have equal pressures, as well as
equal chemical potential for each species. These 3 constraints on 4
unknowns leads to a locus of fluid (crystal) states in the phase
diagram which can coexist with a crystal (fluid) phase. We now have
two lever rules, one for each species, which supply the extra
constraints needed to fix coexisting points, and to fix the volume of
the crystal phase.

In the case of an {\it n}-component system, there are $n+1$
constraints arising from equality of pressure and $n$ chemical
potentials. The problem of finding two coexisting points in the
$n$-dimensional space of the phase diagram is a $2n$-dimensional
problem, leaving us with $(n-1)$-dimensional phase boundaries. There
are now $n$ lever rules, one for each species, allowing us to fix
coexisting points on the phase boundaries, as well as the ratio of
phase volumes. We note that, in the continuously polydisperse case,
the equations remain closed despite the infinite number of
thermodynamic variables. Therefore, in principle at least, we know how
to calculate the equilibrium phase behaviour of any hard sphere
system. The practicalities of phase separation in real systems are a
different matter, which we now proceed to discuss.

\section{Colloidal Systems}

In a polydisperse hard-sphere system each particle has a slightly
different size. In a given sample then, there will be a
thermodynamically large number of species present, and thus the free
energy is a function of the same number of concentration variables.
As such a system tends towards equilibrium, it must minimise its free
energy with respect to these variables. The minimisation in this case
is obviously much more complicated than in the monodisperse situation,
where there is only one relevant parameter.

To reach the equilibrium state from some initial state which an
experimenter prepares by mixing and homogenising, the system must
separate itself into distinct regions of coexisting phases.  Even in
the relatively simple case of monodisperse phase separation, there are
many pathways by which the phase separation may occur
\cite{kinetic}. In the polydisperse case, the situation is complicated
further, as the densities of individual species need not relax at the
same rate as the overall density. One could conceive a situation
whereby phase equilibrium is approached by a two-stage process, as has
been suggested previously for polymeric systems \cite{Warren99}.  In
such a scenario, the polydisperse system initially lowers its free
energy by a quick and expedient demixing of material to form the
separate phase regions. Subsequently, the distributions of species
within the phases are further optimised to attain the absolute minimum
of free energy, which requires particles to be exchanged between the
separate regions. Alternatively, it would be possible for the system
to separate into the two phases with the optimum particle distribution
right from the start. Perhaps more realistically, the separation could
proceed along a path somewhere between these two extremes.

A clue to how the separation does occur is found in experiments on
colloids of attractive particles. Colloidal systems may, given a
sufficiently long ranged inter-particle attraction, exhibit two fluid
phases, analogous to the liquid and gas phases of atomic systems.
Observations of the fluid-fluid phase separation reveal that the
process is approximately as swift in the polydisperse case as in the
near-monodisperse \cite{Davesthesis}.  Given that minimisation of the
polydisperse free energy must be carried out with respect to many more
variables, it might be expected that the additional `sorting' required
would result in a slower relaxation to equilibrium. This suggests that
the initial phase separation is not the end of the story --- the
systems are relaxing to equilibrium by a two-stage process, with an
initial fast separation of material, followed by particle exchange to
absolutely minimise the free energy.

Such an optimisation stage is feasible between two fluid phases as,
given sufficient time, particles can diffuse between phases. In a
hard-sphere fluid-crystal phase separation, however, the system has a
problem. The caging of particles by their neighbours suppresses
diffusion within the crystal to a very low rate, mediated only by
lattice defects. Optimising the population of particles in the
crystal, by exchanging a macroscopic amount of material with the
adjacent fluid region, is therefore unachievable on experimental time
scales. Hence we expect the crystalline population to be arrested in a
non-equilibrium state, so that the footprint of the initial, expedient
stage of phase separation remains observable at late times.

 Therefore, if we wish to make predictions as to the phase behaviour
of hard-sphere colloids, we need to move away from equilibrium
predictions and to model the process of crystal growth. We now develop
such a model, with a view to predicting the size distribution which
forms within a polydisperse hard-sphere crystal.

\section{Crystal Growth Process}
\label{GenDiffusion}

\begin{figure}
  \epsfxsize=8cm \begin{center} \leavevmode \epsffile{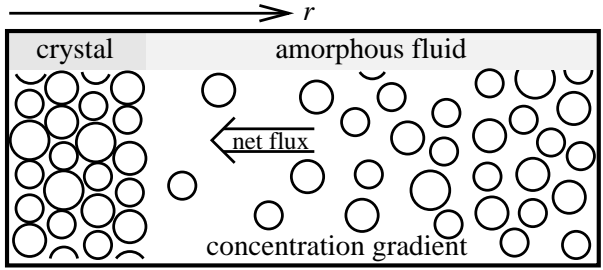}
  \caption{Schematic illustration of the crystal growth mechanism.}
  \label{crystalformation} \end{center}
\end{figure}

In this section, we outline the treatment of the kinetic process
during crystal-fluid phase separation.  We begin by considering a
size-polydisperse sample of colloidal particles suspended in solvent,
prepared with a composition lying in the fluid-crystal coexistence
region of the phase diagram. The sample is assumed to be initially
homogeneous. Eventually, a crystal will randomly nucleate somewhere in
the fluid, creating a concentration gradient in the fluid, as the
immediate vicinity of the growing crystal is depleted of
particles. Particles will diffuse down this gradient towards the
crystal, with the smaller particles diffusing more quickly. The
situation is illustrated in Fig.~\ref{crystalformation}. It should be
noted that the latent heat released during the crystal formation does
not hinder crystallisation. This is because the solvent acts as a heat
bath, keeping the temperature of the system constant.

We wish to find the composition of the crystal, for which we require
the time- and position-dependent composition of the surrounding
fluid. Let us characterise each particle species by its radius,
$a$. The number density $\rho(a,\mbox{\boldmath{$r$}},t)\,\mbox{d}a$
is a function of position and time, as is the diffusive flux
$\mbox{\boldmath{$j$}}(a,\mbox{\boldmath{$r$}},t)\,\mbox{d}a$, for
particles with sizes in the interval $a\to a+\mbox{d}a$.  We begin
with an expression of continuity,
\begin{equation} 
\label{cont1}
  \frac{\partial \rho(a)}{\partial t} = -
  \mbox{\boldmath$\nabla$}\cdot\mbox{\boldmath$j$}(a)
\end{equation}
and a generalisation of Fick's Law to polydisperse particles, to
describe the diffusive dynamics of the fluid phase,
\begin{equation} 
\label{Fick1}
  \mbox{\boldmath$j$}(a,\mbox{\boldmath$r$},t) = -\int
	D(a,a',[\rho(a)])\, \mbox{\boldmath$\nabla$}
	\rho(a',\mbox{\boldmath$r$},t)\: \mbox{d}a'
\end{equation}
which tells us that the flux of a particular species depends on the
concentration gradients of {\em all} species. The diffusion
coefficient $D(a,a',[\rho(a)])$, is a functional of the composition
$\rho(a)$. Later, we will introduce a particular form for $D$, but for
now we continue with the general case.

We must also describe the physics of the interface between the fluid
and crystalline regions. Firstly, we demand conservation of material.
If $r_0$ is the position of the interface, in unit time the interface
advances a distance $\dot{r}_0$. In this time, unit area of the
crystal surface `swallows up' a volume $\dot{r}_0$ of the fluid,
requiring an additional $\dot{r}_0\Delta\rho$ particles, where
$\Delta\rho$ is the (positive) difference in densities across the
interface. So $\dot{r}_0$ is related to the flux across the interface
by
\begin{equation}
\label{fluxtrick}
	\mbox{\boldmath$j$}(a)
	= -\mbox{\boldmath$\hat{n}$}\,\dot{r}_0\,\Delta\rho(a) 
\end{equation}
where {\boldmath$\hat{n}$} is a unit normal to the interface.

Various empirical approximations exist for the flux across the
interface between non-equilibrium phases. In the following, the
details of the approximation will turn out to be unimportant. We
choose the standard Wilson-Frenkel law \cite{wilson-frenkel} which,
for a monodisperse system, is given by
\begin{displaymath}
	\mbox{\boldmath$j$} = -\mbox{\boldmath$\hat{n}$}\nu D_s
	\frac{(\exp\Delta\mu)-1}{\Lambda},
\end{displaymath}
where $\nu$ is a constant, $D_s$ is a short-time self-diffusion
coefficient, $\Delta\mu$ is the chemical potential difference across
the interface (with units such that $k_B T\equiv 1$) and $\Lambda$ is
the interfacial width. We generalise it for our polydisperse system to
\begin{equation}
\label{WF} 
	\mbox{\boldmath$j$}(a)=\frac{-\mbox{\boldmath$\hat{n}$}}{\Lambda} 
	\int \Gamma(a,a',[\rho(a)])\; \{\exp{\Delta\mu(a')}-1\}\,\mbox{d}a'
\end{equation}
where $\Gamma(a,a',[\rho(a)])$ is a mobility for particles of radius
$a$ due to a chemical potential difference in particles of radius
$a'$. The mobility is a functional of the whole set of concentrations
present.

We now search for a solution to our equations. For simplicity, let us
assume that the crystal grows with spherical symmetry, neglecting the
possibilities of faceting or dendritic growth. We try a solution in
which lengths scale as $t^{1/2}$, the validity of which will be
examined {\it a posteriori}. We may then replace $r\to R\, t^{1/2}$.
With this assumption, we can write the density of particles of radius
$a$ as $\rho(a,r,t)=\rho(a,R)$. At a given value of R, the density
remains constant in time; the diffusive flux, however, will fall off
as $t^{-1/2}$, as it is proportional to density gradients.  We write
the flux as
$\mbox{\boldmath$j$}(a,r,t)=t^{-1/2}\mbox{\boldmath$J$}(a,R)$.  Let us
use these variables to re-write the relevant equations. The
generalised Wilson-Frenkel law becomes
\begin{displaymath} 
	\mbox{\boldmath$J$}(a,R)=-\mbox{\boldmath$\hat{n}$} \int
    	\Gamma(a,a',[\rho(a)]) \left\{\frac{\exp{\Delta
    	\mu(a')}-1}{\Lambda t^{-\frac{1}{2}}}\right\}\mbox{d}a'.
\end{displaymath}
Note that $\Lambda$, the width of the interface, is the only length
which does not scale as $t^{1/2}$.  Since the flux {\boldmath$J$} is
now independent of $t$, the RHS must also be time independent, so we
have
\begin{displaymath} 
	\exp{\Delta \mu(a)}-1 = g(a)\,t^{-\frac{1}{2}}
\end{displaymath}
with some (unknown) function $g(a)$, which gives
\begin{equation} 
\label{Deltamu} 
	\Delta\mu(a) \to g(a)\,t^{-\frac{1}{2}} \to 0
\end{equation}
at late times, confirming our expectation that the interface tends to
local equilibrium. The Wilson-Frenkel law has demonstrated the
relevant physics, that lengths controlling sub-interfacial dynamics
remain microscopic and therefore become irrelevant.

The equation of continuity may also be written in terms of the scaled
variables. As spherical symmetry is assumed, we transform to
$d$-dimensional spherical coordinates. Hence Eq.~(\ref{cont1}) becomes
\begin{equation} 
\label{continuity}
	\frac{\partial \rho (a,R)}{\partial
    	R}=\frac{2}{R}\left(\frac{\partial}{\partial
    	R}+\frac{d-1}{R}\right)J(a,R).
\end{equation}
Similarly, we write Fick's law [Eq.~(\ref{Fick1})] as,
\begin{equation} 
\label{Fick}
	J(a,R) = -\int D(a,a',[\rho(a)]) \frac{\partial
      	\rho(a',R)}{\partial R}\,\mbox{d}a'.
\end{equation}
From Eq.~(\ref{fluxtrick}), we obtain
\begin{equation} 
\label{continterface}
	J(a,R_0) = -\frac{1}{2}R_0[\rho_c(a)-\rho(a,R_0)]
\end{equation}
where $\rho_c(a)$ is the density distribution in the crystal and $R_0$
is the value of $R$ at the interface. Note that $\rho_c(a)$ is
independent of position and time. Our scaling solution yields a unique
answer (though not the equilibrium one) for the crystalline
composition (for a given system and parent). In other words, the
crystal grows uniformly.

We now have a closed pair of time-independent equations
[Eqs.~(\ref{continuity}, \ref{Fick})], with the boundary condition at
$R=R_0$ given by Eq.~(\ref{continterface}) and, from Eq.~(\ref{Deltamu}),
local thermodynamic equilibrium across the interface. The boundary
condition at $R\to\infty$ is $\rho(a,R)\to\rho_p(a)$ since the {\em
parent} distribution $\rho_p(a)$ exists in the bulk. The fact that
this asymptote remains time-independent is a consequence of our
scaling solution, for which all lengths go as $t^{1/2}$. We see, then,
that this scaling solution describes the regime of growth {\em before}
the distant composition of the super-saturated fluid has been depleted
(i.e.~while condensation nuclei grow as if isolated), but {\em after}
the crystal-fluid interface has locally equilibrated following the
initial transients of nucleation.

\section{A Modified Lever Rule}
\label{ModLever}

In order that an analytic solution is possible, we now make the
simplest possible choice of the diffusion coefficient present in
Fick's law, and write
\begin{equation}
	D(a,a') = D_0(a)\,\delta(a-a') = \frac{kT}{2d\pi\eta
	a}\,\delta(a-a'),
\end{equation}
where $D_0(a)$ is the monodisperse diffusion coefficient of a particle
of size $a$, taken to be of Stokes-Einstein form.  In making this
approximation, our generalised Fick's law reduces to the ordinary form
for each species independently. This neglects the flux of a given
species due to density gradients of all other species. This becomes
correct in the low density limit \cite{Batchelor}, but will be
incorrect at the large densities required for coexistence in a
hard-sphere system. Here, the generalised Fick's law has been
discarded, but a more sophisticated treatment could utilise it along
with a more accurate approximation to the diffusion coefficient
\cite{Batchelor}.

We now wish to solve our system of equations. We begin with
Eq.~(\ref{Fick})
\begin{equation} 
\label{fick} 
	-J(a,R)=D_0 (a)\frac{\partial \rho(a,R)}{\partial R}
\end{equation}
Substituting in Eq.~(\ref{continuity}), and with a little work, we
find
\[ 
	J(a,R) = J_0(a) \left(\frac{R}{2\sqrt{D_0(a)}}\right)^{(1-d)}
	\exp\left({-\frac{R^2}{4D_0(a)}}\right)
\]
with $J_0(a)$ appropriately defined. Now using the expression of
continuity of particles at the interface (\ref{continterface}), we fix
$J_0(a)$, leaving us with
\begin{equation}
\label{J}
    J(a,R) = -\frac{1}{2}R_0[\rho_c(a)-\rho_f(a)]\frac{R_0}{R}^{d-1}
    \exp{\left(\frac{R_0^2-R^2}{4D_0(a)}\right)},
\end{equation}
where the density distribution in the fluid at the interface has been
written as $\rho_f(a)$. Integrating Eq.~(\ref{fick}), and applying the
boundary condition at $R\to\infty$, we obtain for $\rho(a,R)$,
\[
	\rho(a,R) = \rho_p(a)+\int^\infty_R
     	\frac{J(a,R')}{D_0(a)}\,\mbox{d}R'.
\]
On substitution from Eq.~(\ref{J}), we obtain
\begin{equation}
\label{important}
    	\frac{\rho_p(a)-\rho(a,R)}{\rho_c(a)-\rho_f(a)} =
	\frac{R_0}{2D_0(a)}\int^\infty_R
	\left(\frac{R_0}{R'}\right)^{d-1}\exp{\left(
	\frac{R_0^2-R^{\prime 2}}{4D_0(a)}\right)}\, \mbox{d}R'.
\end{equation}
This yields a closed form expression for $\rho(a,R)$. For our present
purposes, we evaluate Eq.~(\ref{important}) at $R=R_0$. With a change of
variables to $u\equiv R'/R_0$, we find
\begin{equation}
\label{FLR}
	\chi(a) \equiv \frac{\rho_p(a)-\rho_f(a)}{\rho_c(a)-\rho_f(a)}
	= f_d\left(\frac{R_0}{2\sqrt{D_0(a)}}\right)
\end{equation}
where
\begin{equation} 
	f_d(x) \equiv 2 e^{x^2} x^d
	\int^\infty_x\frac{\exp(-u^2)}{u^{d-1}}\,\mbox{d}u = \left\{
	\begin{array}{ll} \sqrt{\pi}\,x\,e^{x^2}\,\mbox{erfc}\,x\; &
	\mbox{ for $d=1$} \\ 2x^2 ( 1 -
	\sqrt{\pi}\,x\,e^{x^2}\,\mbox{erfc}\,x )\; & \mbox{ for
	$d=3$}.  \end{array} \right.
\end{equation}
This is an important result in our treatment of the kinetics.  Recall
that, in a system {\em at equilibrium}, conservation of material is
expressed by the lever rule [Eq.~(\ref{lever})], which may be
re-expressed as
\begin{equation}
\label{LR}
    \frac{\rho_p(a)-\rho_f(a)}{\rho_c(a)-\rho_f(a)}=\frac{V_c}{V},
\end{equation}
where $V$ and $ V_c$ are the overall system volume and the volume of
the crystalline phase respectively. The densities have the same
meanings as in Eq.~(\ref{FLR}), with the exception that here
$\rho_f(a)$ refers to the equilibrium fluid, which will be the same
throughout that phase, whilst in the non-equilibrium case the
subscript $f$ refers only to the fluid at the interface. We stress
that the lever rule does {\em not} hold in the situation under
consideration. Rather, on comparing Eqs.~(\ref{LR}) and (\ref{FLR}), we
note that Eq.~(\ref{FLR}) may be considered an alternative to the lever
rule in our non-equilibrium system. Obviously we must conserve matter
in the non-equilibrium system as a whole, but here we are only
considering a certain region, into which there is a flux of material.

In the same way as the lever rule closes the set of equations
governing equilibrium phase behaviour (and fixes $V_c/V$),
Eq.~(\ref{FLR}) closes the same set of equations in the non-equilibrium
case (and fixes $R_0$). As we are not conserving material in this
case, the tie lines predicted in the non-equilibrium phase diagram,
using Eq.~(\ref{FLR}), need not be straight. In summary, the crystal
formed has a composition that lies on the equilibrium phase boundary,
since it coexists with a local region of fluid, but this composition
appears at the end of the `wrong' tie-line.

So now we have an alternative to the lever rule for our
non-equilibrium system. Using this, along with conditions of local
equilibrium at the crystal fluid interface, we shall predict the phase
behaviour of our polydisperse system.

\section{Local Equilibrium}
\label{Perturb}

In order to solve Eq.~(\ref{Deltamu}) for the locally coexisting density
distributions that ensure chemical (and mechanical) balance across the
crystal-fluid interface, we need a technique by which to calculate
phase equilibria in a polydisperse system. This is a difficult task,
even in the simplest case of an equilibrium system for which the free
energy is known, since the free energy is a function(al) of an
infinity of concentration variables. Several different approaches have
been developed to tackle the problem
\cite{polypapers}, most relying on a numerical stage to the
solution. In pursuit of a concise result, we adopt a perturbative
approach which yields analytic answers in a closed form. The method
has previously been applied to equilibrium problems
\cite{Evans98,Evans2001}. We shall adapt the mathematics to deal with
the non-equilibrium aspects of the system, by replacing the lever rule
with our non-equilibrium equivalent.

We take a monodisperse system as a reference state, and treat the
polydispersity as a perturbation.  This permits us to deal with
systems in which the degree of polydispersity is, in some sense,
small, and will only be applicable if the polydisperse system behaves
similarly to the monodisperse limit. In particular, we expect to find
coexistences corresponding to those present in the monodisperse
system, with binodal concentrations and other properties altered a
little by the polydispersity. For details of the method, the reader is
directed to Ref.~\cite{Evans2001}.

Let us briefly explain the notation used. For a species of particles
of radius $a$, we define a small, dimensionless number $\epsilon
\equiv (a-a_0)/a_0$, to quantify its deviation from the radius $a_0$
of particles in the monodisperse reference system.  We denote the
number density of particles of `size' $\epsilon$ as
$\rho(\epsilon)\,\mbox{d}\epsilon$. Normalised distributions are
denoted $p(\epsilon)$, and subscripts label the relevant population of
particles, be it that of the crystal ($c$), fluid ($f$), or parent
($p$). Angled brackets denote averages over the relevant
distribution. We choose the reference size $a_0$ to be the mean of our
parent distribution so that, by definition, $\langle\epsilon\rangle_p$
is identically zero.

Our approach is as follows. We wish to find the conditions for
equality of pressure and chemical potentials across the interface
between crystal and fluid. The pressure of a phase is a functional of
the whole distribution of densities within it. Likewise, the chemical
potential of a species is a functional of the density distribution,
and is also a function of the particular species in question,
characterised by its size deviation $\epsilon$. So we need to solve
the infinity (plus one) of equations
\begin{eqnarray}
	P[\rho_f(\epsilon)] &=& P[\rho_c(\epsilon)]
\label{pressure} \\
	\mu(\epsilon, [\rho_f(\epsilon')]) &=& \mu(\epsilon,
	[\rho_c(\epsilon')]) \;\; \forall \: \epsilon,
\label{chempotl}
\end{eqnarray}
the latter of which is the late-time limit of Eq.~(\ref{Deltamu}).
These relations between $\rho_f(\epsilon)$ and $\rho_c(\epsilon)$
define the phase boundaries of an equilibrium phase diagram in
$\rho(\epsilon)$-space. Simultaneously applying the modified lever
rule [Eq.~(\ref{FLR})] then fixes uniquely the coexisting distributions.
To solve Eqs.~(\ref{pressure}, \ref{chempotl}), we expand their slowly
varying parts in $\epsilon$. At zeroth order, this yields the
conditions for local equilibrium in the monodisperse system, for which
solutions are known. A problem exists in expanding
Eq.~(\ref{chempotl}). In the reference system, species for which
$\epsilon \neq 0$ are unpopulated, so their chemical potential is
negative infinity. This singularity is logarithmic in the density, as
is the case for an ideal gas, $\mu^{\rm id}(\epsilon) =
\ln\rho(\epsilon)$. To circumvent the problem, we shall subtract off
this singular part, and work with the excess chemical potential
$\mu^{\rm ex}\equiv\mu-\mu^{\rm id}$, in terms of which,
Eq.~(\ref{chempotl}) becomes
\begin{equation}
\label{eqchempot}
	\rho_f(\epsilon)\, \exp\mu^{\rm ex}_f(\epsilon,
	[\rho_f(\epsilon)]) = \rho_c(\epsilon)\, \exp\mu^{\rm ex}_c
	(\epsilon, [\rho_c(\epsilon)])
\end{equation}
which may be substituted into Eq.~(\ref{FLR}), the modified lever rule,
to yield
\begin{equation}
\label{compact}
	\rho_c(\epsilon) = \frac{\rho_p(\epsilon)}{\chi(\epsilon) +
	\big(1-\chi(\epsilon)\big) \exp[\mu^{\rm
	ex}(\epsilon,[\rho(\epsilon)])]^c_f}
\end{equation}
where the notation $[x]^c_f$ denotes the difference in quantity $x$
between the phases, and we have written $\chi(\epsilon)$ for
$\chi((1+\epsilon)a_0)$ as defined in Eq.~(\ref{FLR}).

If we now express each density distribution $\rho(\epsilon)$ in terms
of its normalisation $\rho$ and moments $\{\langle\epsilon\rangle,
\langle\epsilon^2\rangle, \ldots\}$ then the excess chemical
potential, appearing on the R.H.S. of Eq.~(\ref{compact}) can be
expanded as \cite{Evans2001}
\begin{equation}
\label{excesschempot}
	\mu^{\rm ex}(\epsilon,[\rho(\epsilon)]) = \mu^{\rm
	ex}_0(\rho,\langle\epsilon\rangle) +
	\frac{A(\rho)}{\rho}\,\epsilon + {\cal O}(\epsilon^2)
\end{equation}
in terms of its value $\mu^{\rm ex}_0$ for the mean species in the
phase (which differs a little \cite{Evans2001} from the overall mean
at $\epsilon=0$), and the function $A(\rho)$ which parameterises the
variation of excess chemical potential with size. Note that
$A(\rho)=\rho\,\mbox{d}\mu^{\rm ex}/\mbox{d}\epsilon$, with the
derivative evaluated in the limit of a narrow distribution.

Expanding the exponential in Eq.(\ref{compact}), and writing
$\chi(\epsilon)=\chi_0+\epsilon\,\chi_1 + O(\epsilon^2)$, we obtain
\begin{equation}
\label{crystal.density}
	\rho_c(\epsilon) =
	\frac{\rho_p(\epsilon)\:(\xi+1)}{1+\chi_0\,\xi} \left\{ 1 -
	\epsilon\, \frac{\chi_1 \,\xi + (1-\chi_0)
	[A/\rho]^c_f}{1+\chi_0\,\xi} + O(\epsilon^2)\right\}
\end{equation}
where $\xi\equiv\exp(-[\mu^{\rm ex}_0]^c_f)-1$, with
$\rho_f(\epsilon)$ given, from Eq.~(\ref{FLR}), by
$(\rho_p-\rho_c\,\chi)/(1-\chi)$.  The overall density in either phase
is then given by integration over all sizes.  This allows one to
obtain an expression for {\it normalised} size distributions.  The
difference in this distribution between phases is thus found to obey
an expression that is independent of kinetic parameters
\begin{equation}
\label{result}
	[p(\epsilon)]^c_f= -p_p(\epsilon)\left\{\epsilon\,
	[A/\rho]^\alpha_\beta + O(\epsilon^2)\right\}.
\end{equation}
Combining this result with the modified lever rule [Eq.~(\ref{FLR})]
recovers expressions for the normalised size distributions either side
of the crystal-fluid interface
\begin{eqnarray}
  p_c(\epsilon) &=& p_p(\epsilon) \left( 1 +
	\left\{(\chi_0-1)\frac{\rho_f}{\rho_p} [A/\rho]^c_f - \chi_1
	\frac{(\rho_c-\rho_f)}{\rho_p}\right\}\epsilon +
	O(\epsilon^2)\right)
\label{normdistns1}	\\ 
  	p_f(\epsilon) &=& p_p(\epsilon)\left( 1+
	\left\{\chi_0\frac{\rho_c}{\rho_p} [A/\rho]^c_f - \chi_1
	\frac{(\rho_c-\rho_f)}{\rho_p}\right\}\epsilon +
	O(\epsilon^2)\right).
\label{normdistns2}
\end{eqnarray} 
Note that Eq.~(\ref{pressure}) for equality of pressure was not
required. It affects the results only at higher order in $\epsilon$.

In order to employ these equations, we need expressions for $\chi_0$,
$\chi_1$ (which depend on $D_0(\epsilon)$ and the unknown $R_0$), and
the densities of each phase in the non-equilibrium polydisperse
system.  Consider the behaviour of $R_0$, the scaled position of the
interface. This controls the rate of growth, via $r_0=R_0\, t^{1/2}$.
Physically, $R_0$ must depend on the composition of the system,
i.e.~on the density distribution $\rho_p(\epsilon)$, so
$R_0=R_0(\rho_p, \langle\epsilon\rangle_p, \langle\epsilon^2\rangle_p,
\ldots)$.  Assuming that $R_0$ can be smoothly expanded about its
monodisperse value thus:
\[
	R_0 = R_{\rm m}(\rho_p) +R_1(\rho_p)\,\langle\epsilon\rangle_p
	+R_{11}(\rho_p)\,\langle\epsilon\rangle_p^2
	+R_2(\rho_p)\,\langle\epsilon^2\rangle_p +\ldots
\]
we find that, to first order, it is unchanged from the monodisperse
value, since $\langle\epsilon\rangle_p \equiv 0$. We may therefore use
the monodisperse value of $R_0$ in our first order calculation. Hence,
we can take $\chi_0$ to be the monodisperse value of
$f_d\left(R_0/2\surd D_0(\epsilon)\right)$, given by
Eq.~(\ref{FLR}). That is,
\begin{equation}
	\chi_0 = \frac{\rho_p-\rho_f^{\rm m}} {\rho_c^{\rm
		m}-\rho_f^{\rm m}}
\end{equation}
i.e.~$\chi_0$ is the distance of the quench from the monodisperse
phase boundary, as a fraction of the width of the coexistence
region. The unknown crystal growth rate $R_0$ is then given, in terms
of the inverse of the function $f_d$, by
\begin{equation}
	R_0 = 2\sqrt{D_0}\: f_d^{-1}(\chi_0) \; +
O(\langle\epsilon^2\rangle_p)
\end{equation}
with $D_0$ evaluated at $\epsilon=0$. Taylor expansion of
Eq.~(\ref{FLR}) then gives
\begin{displaymath}
	\chi_1 = \frac{\mbox{d}\chi(\epsilon)}{\mbox{d}\epsilon} =
	-\frac{R_0\, D_0'}{4D_0^{3/2}}\,
	f'_d\left(\frac{R_0}{2\sqrt{D_0}}\right)
\end{displaymath}
where $D_0'=\mbox{d}D_0/\mbox{d}\epsilon$ evaluated at $\epsilon=0$.

By integration of Eq.~(\ref{crystal.density}), one finds that (to first
order) the overall densities of each phase are unchanged from the
monodisperse values. Hence, on the RHS of Eqs.~(\ref{normdistns1}) and
(\ref{normdistns2}), we may use the monodisperse, equilibrium values of
these quantities.

\section{Results}
\label{results}

The formal expressions obtained are now applied to the case study of
non-attracting polydisperse hard spheres, for which the value of the
parameter $[A/\rho]^c_f$ has been determined previously
\cite{Evans2001,Bolhuis96} as $-3.55$ in units of $kT$.

\begin{figure}
  \begin{center} \epsfxsize=8.5cm \leavevmode
  \epsffile{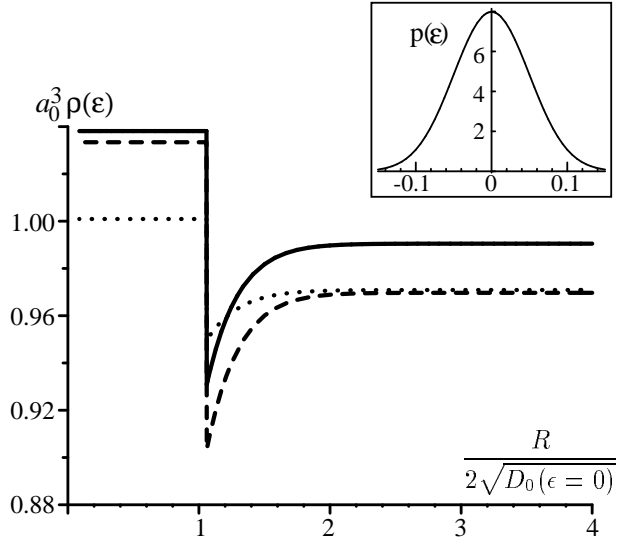} \caption{Density profiles for three
  of the infinite number of species in the hard-sphere system: the
  mean sized species ($\epsilon=0$, solid line), a large species
  ($\epsilon=0.01$, dashed) and a small species ($\epsilon=-0.01$,
  dotted). The normalised parent distribution is Gaussian (inset),
  with standard deviation $\surd\langle\epsilon^2\rangle=0.05$; the
  parent concentration is $\phi_p=0.52$.}  \label{profiles}
  \end{center}
\end{figure}

Equations (\ref{important}), (\ref{normdistns1}) and (\ref{normdistns2})
allow calculation of density profiles in our system. An example is
plotted in Fig.~\ref{profiles}, which shows density profiles
calculated for a non-attractive hard-sphere system in $d=3$ spatial
dimensions. One can see that the diffusion process transports small
particles to the crystal-fluid interface in relatively greater
abundance than it does large particles (relative to the numbers in the
parent distribution). Thus, when thermodynamic effects (of local
equilibrium at the crystal-fluid interface) decide which particles are
incorporated into the crystal, the choice is made from a biased
version of the parent. The result of this is that the crystal is made
up of particles which are, on average, smaller than would be the case
in equilibrium.

\begin{figure} 
  \epsfxsize=8cm \begin{center} \leavevmode \epsffile{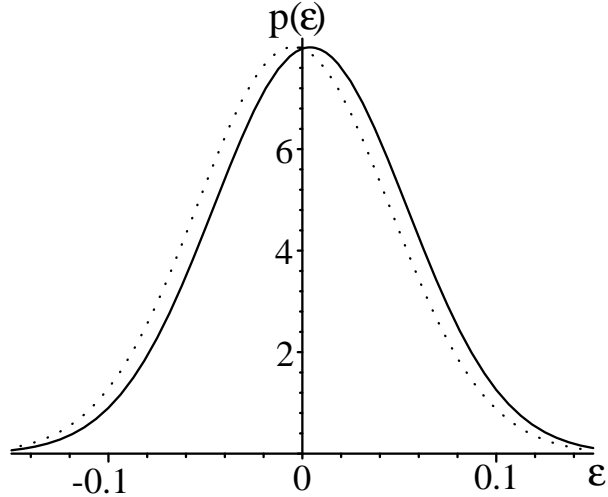}
  \caption{The {\em equilibrium} size distribution in the crystal
  (solid line) and fluid (dotted).}  \label{eq.distn} \end{center}
\end{figure}

\begin{figure}
  \epsfxsize=6.5cm \begin{center} \leavevmode
  \epsffile{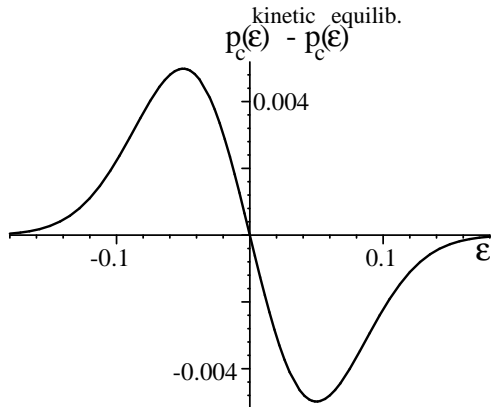} \caption{The difference in normalised size
  distributions in the crystal between that calculated using the
  present kinetic model, and that of the equilibrium theory (for the
  parent used in Fig.~\protect\ref{profiles}).}
\label{difference}
  \end{center}
\end{figure}

\begin{figure}
  \epsfxsize=8cm \begin{center} \leavevmode \epsffile{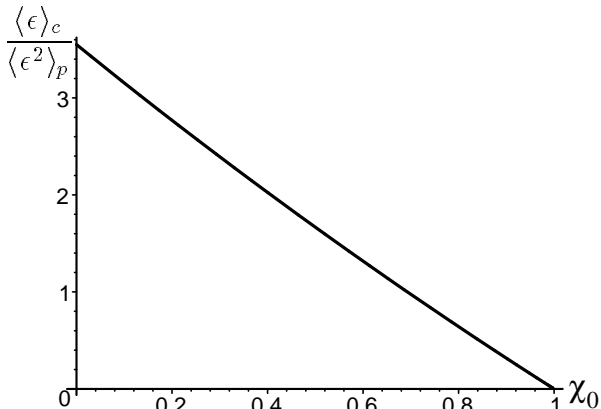}
  \caption{The mean size deviation of particles in the crystal, in
  units of the overall variance,
  $\langle\epsilon\rangle_c/\langle\epsilon^2\rangle_p$, as a function
  of relative supersaturation of the initial fluid phase, $\chi_0$.}
  \label{crystalmean} \end{center}
\end{figure}

Equation (\ref{normdistns1}) allows us to calculate the size
distribution in the crystal, given any narrow parent. As an example we
use a Gaussian parent. For comparison the resulting distributions in
the fluid and crystal, for a system which has reached thermodynamic
equilibrium (as predicted by the equilibrium perturbation theory
\cite{Evans2001}) are displayed in Fig.~\ref{eq.distn}. We note that
the crystal has a preference for larger particles, telling us that
entropy is maximised if the particles are partitioned in this
way. Heuristically, this is because the particles' positional entropy
in the fluid phase is increased if more space is made available by
removing the larger particles to the crystal phase. What then is the
effect of the growth process on the size distributions? The effect can
be seen by plotting the difference between size distributions in the
crystal as predicted by the kinetic and equilibrium calculations. This
is shown in Fig.~\ref{difference}.
We see the effect of kinetics is to bias the distribution in the
crystal towards smaller sizes. We take the first moment of
Eq.~(\ref{normdistns1}), to find the mean size
$\langle\epsilon\rangle_c$ in the crystal, which is a function of the
overall concentration in the system. It is plotted in
Fig.~\ref{crystalmean} as a function of the relative supersaturation
$\chi_0$ beyond the fluid phase boundary.
At $\chi_0=0$, the system is at equilibrium, at its cloud point, since
the fluid is not supersaturated, and the crystal phase at coexistence
occupies an infinitesimal volume. As a result, the growth rate $R_0$
vanishes, and the mean size deviation in the crystal attains its
equilibrium value
$\langle\epsilon\rangle_c=3.55\langle\epsilon^2\rangle_p$. As the
supersaturation of the initial fluid state increases, the mean
particle size in the crystal becomes smaller, in an almost (though not
quite) linear fashion. In the limit where the initial state is so
dense that crystallisation induces no density change, the mean size in
the crystal equals that in the parent ($\langle\epsilon\rangle=0$).

Note that, in the hard-sphere system, though the effect of kinetics is
to reduce the mean particle size in the crystal, it remains larger
than the overall parental mean, i.e.~$\langle\epsilon\rangle_c \geq
0$.  Hence, in a sense, the thermodynamic effects of crystal-fluid
partitioning win over the kinetics. We find this is a consequence of
the small density change at the hard-sphere crystal-fluid phase
transition ($\rho_c/\rho_f-1\approx 10\%$). In a second case study
(not presented here), we have substituted into Eq.~(\ref{normdistns1})
parameters appropriate to a system of hard spheres with attractive
interactions \cite{Evans2001} (in particular, the `depletion'
interaction arising in colloid-polymer mixtures
\cite{lekkerkerker}). In that case, the coexistence region can become
much broader since, with strong attractions, even a very dilute gas of
hard spheres can crystallise. Then, we find a range of values of
$\chi_0$ for which crystals form from particles {\em smaller} on
average than the mean composition of the system,
i.e.~$\langle\epsilon\rangle_c<0$, signalling the dominance of the
kinetic effects presented here.

\section{Conclusion}
\label{conc}

The kinetics of polydisperse systems are sufficiently complex that few
analytical or even numerical studies have been attempted.  Exceptions
are Refs.~\cite{Warren99} and \cite{Clarke2001}.

We have argued that colloidal hard-sphere systems (which are
inevitably polydisperse) form crystals whose composition is not at
equilibrium, i.e.~does not maximise the entropy of the two-phase
ensemble. Instead, particles are caged in the crystalline
structure. In practise, some small-scale re-arrangements within the
crystal can take place in the presence of lattice defects, but the
rate of particle diffusion is negligible (particularly with
inter-particle attractions) compared to that in the fluid phase. As a
result, the distribution of particle sizes frozen into a colloidal
crystal remains as a relic of its growth mechanism.

The establishment of chemical equilibrium requires a significant
fraction of each chemical species to be exchanged many times between
the crystal and fluid phases. A hypothetical system in which this
occurs, so that phase space is fully explored and distributions are
optimised, would preferentially partition more large particles into
the crystal phase, so that particles in the coexisting fluid have more
space in which to enjoy positional entropy. In contrast, the diffusive
growth process biases the crystalline composition towards small
particles, since they can travel most quickly from the distant bulk of
the ambient fluid. While the mean particle size at the crystal-fluid
interface must be larger in the crystal side than in the adjacent
fluid, the largest, most sluggish particles remain predominantly in
the fluid bulk.

The regime for which we were able to find analytical solutions of the
equations of motion was at intermediate times, when all relevant
lengths scale with time as $t^{1/2}$. That is, after transients
associated with the initial random nucleation event have passed so
that, on the scale of the interfacial width (a few particle
diameters), the phases are at local equilibrium, but before the fluid
zone of depleted concentration around each condensation nucleus begins
to overlap with its neighbours, so that the distant fluid composition
still asymptotes to the initial (`parent') mixture. Once these
depleted regions do significantly overlap, the supersaturation of the
fluid phase is soon exhausted. The remaining mix of predominantly
large particles, which have remained in the fluid, will eventually
coat each crystallite, and growth will cease when no concentration
gradient remains in the fluid. Hence, in laboratory samples, we expect
the inert colloidal crystallites to have an unknown core of a few
particles which collected together during the early nucleation stage,
a significant bulk of the uniform composition we have calculated, and
a non-uniform coating of larger particles. Happily, equilibrium phase
diagrams, which have previously been calculated, are not redundant, as
the uniform composition from the scaling regime of crystal growth does
lie on the equilibrium phase boundary, though not on the tie-line
specified by the usual lever rule. For our calculation to be useful,
the core region of each crystallite, whose composition we do not know,
must be small compared with the whole crystallite, which requires a
low concentration {\em of condensation nuclei} to appear in the
system.

The above scenario should hold, no matter how the diffusion matrix
(which determines the flux of each species induced by concentration
gradients of any other species) varies with concentration. Though
collective motions and many-body interactions lead to a non-linear
diffusion equation for the fluid \cite{Batchelor}, this affects
details of the shape of the concentration profiles in
Fig.~\ref{profiles}, but does not alter the $t^{1/2}$ growth law or
the conditions of local interfacial equilibrium, leading to a uniform
crystalline composition. Also, the qualitative principle remains, that
small particles diffuse most quickly.

To find quantitative solutions, however, we were compelled to make two
approximations. The first, in order to diagonalize the diffusion
matrix, was a low concentration approximation which, though it has a
regime of validity near the fluid phase boundary of attractive
systems, is otherwise quantitatively poor. In its favour, it yields
qualitatively significant results. The trend in
Fig.~\ref{crystalmean}, for instance, is correct, tending as it does
from an equilibrium result at the fluid phase boundary (zero
supersaturation, $\chi_0=0$) to a total absence of demixing at the
crystal boundary ($\chi_0=1$).

Our second approximation holds true for many experimental systems.
Hard-sphere colloids can be synthesised \cite{synthesize} with a
narrow distribution of particle sizes (typically $2-10\%$ tolerance in
the radii), for which perturbation about a monodisperse reference
state, to first order in size deviations, has been shown to yield
accurate results for two-phase equilibrium \cite{Evans98,Evans2001},
which is local in this case. The perturbation expansion holds, given
that the perturbation (i.e.~the width of the distribution) is
sufficiently small. With increasing polydispersity, the system is
expected to exhibit singular behaviour, partitioning its particles
into several coexisting crystal phases of more uniformly sized
particles
\cite{Bartlett98}, or forming alloys with more complex unit cells
\cite{Eldridge95,Bartlett92}, to avoid costly lattice interstitials of
very mis-sized particles. In that case, the growth may be controlled
by the kinetics of segregation at interfaces, leading to different
physics from the diffusion-limited regime calculated here.

\acknowledgments
We thank Michael Cates for constructive discussions.  RMLE
acknowledges the support of the Royal Society of Edinburgh (SOEID
fellowship), and the EPSRC (GR/M29696).

\end{document}